\documentclass[sn-mathphys]{sn-jnl}

\jyear{2022}%

\theoremstyle{thmstyleone}%
\theoremstyle{thmstyletwo}%

\theoremstyle{thmstylethree}%

\begin{document}

\title[M\"ossbauer spectroscopy of $^{227}$Ac]{A new avenue in the search for CP violation: M\"ossbauer spectroscopy of $^{227}$Ac}


\author*[1,2]{\fnm{Marcus} \sur{Scheck}}\email{marcus.scheck@uws.ac.uk}

\author[1,2]{\fnm{Robert} \sur{Chapman}}

\author[3,4]{\fnm{Jacek} \sur{Dobaczewski}}

\author[5]{\fnm{Claude} \sur{Ederer}}

\author[6]{\fnm{Peter} \sur{Ivanov}}

\author[6]{\fnm{Guiseppe} \sur{Lorusso}}

\author[1,2]{\fnm{David} \sur{O'Donnell}}

\author[7]{\fnm{Christian} \sur{Schr\"oder}}

\affil*[1]{\orgdiv{School of Computing, Engineering, and Physical Sciences}, \orgname{University of the West of Scotland}, \orgaddress{\street{High Street}, \city{Paisley}, \postcode{PA11 3PX}, \country{UK}}}

\affil[2]{\orgdiv{SUPA}, \orgname{Scottish Universities Physics Alliance}, \orgaddress{\country{UK}}}

\affil[3]{\orgdiv{Department of Physics}, \orgname{University of York}, \orgaddress{\city{Heslington}, \postcode{YO10 5DD},  \country{UK}}}

\affil[4]{\orgdiv{Department of Physics}, \orgname{University of Warsaw}, \orgaddress{\city{Warsaw}, \postcode{PL-02-093},  \country{Poland}}}

\affil[5]{\orgdiv{Materials Theory}, \orgname{ETH Z\"urich}, \orgaddress{\city{Z\"urich}, \postcode{CH-8093},  \country{Switzerland}}}

\affil[6]{\orgdiv{} \orgname{National Physical Laboratory}, \orgaddress{\city{Teddington}, \postcode{TW11 0LW},  \country{UK}}}

\affil[7]{\orgdiv{Biological and Environmental Sciences}, \orgname{University of Stirling}, \orgaddress{\city{Stirling}, \postcode{FK9 4LA},  \country{UK}}}


\abstract{This work proposes a new avenue in the search for CP-violating odd-electric and even-magnetic nuclear moments. A promising candidate to find such moments in the ground state is the quadrupole-deformed and octupole-correlated nucleus 227-actinium. In this nucleus, the 27.4-keV $E1$~transition that connects the $3/2^+$ parity-doublet partner and the $3/2^-$ ground state is perfectly suited to apply the sensitive technique of recoil-free selfabsorption, commonly known as M\"ossbauer spectroscopy. In this experimental approach, the lifetime of the $3/2^+$ upper parity-doublet partner allows an estimate of the lower limit of $\Delta E = 2\cdot \Gamma_{\gamma}$= $23.7(1) \times 10^{-9}$~eV for the achievable energy resolution to be made. This resolution must be exceeded by the interaction of a CP-violating moment and the corresponding multipole moment of the field distribution in the lattice. This work presents the first ideas for patterns caused by CP-violating moments on the expected quadrupole splitting and nuclear Zeeman effect.}




\maketitle

\section{Introduction}\label{sec1}

The observation of enhanced $B(E3, 0^+\rightarrow 3^-_1)$ excitation strength in several lanthanide \cite{Bucher1,Bucher2} and actinide \cite{Spagnoletti,Butler,Gaffney,Wolle} isotopes indicates octupole correlations in the qua\-drupole-deformed ground state of at least some of these nuclei. The resulting quadrupole-octupole deformed pear shape is predicted to enhance the possible CP-violating laboratory Schiff moment \cite{Spevak,Auerbach,Naftali,Zelevinsky,Dobacewski,Markus,ViktorM2}.

\begin{figure}[t]
\resizebox{1.0\textwidth}{!}{\includegraphics{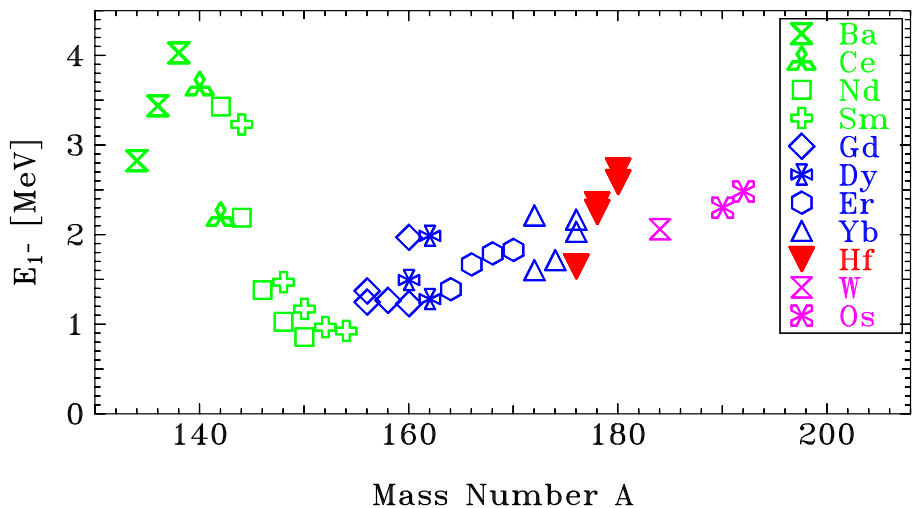}}
\caption{\label{e1-a} Systematics of the excitation energy $E_{1^-}$
of the first excited $1^-_1$ level in the lanthanide/rare-earth
region. Figure is taken from Ref.~\cite{ScheckHf}.}
\end{figure}

\begin{figure}[t]
\resizebox{1.0\textwidth}{!}{\includegraphics{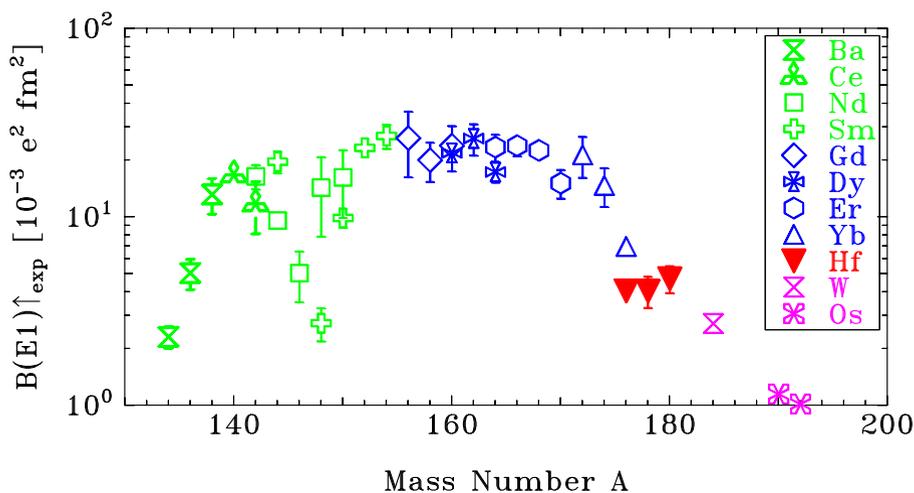}}
\caption{\label{BE1_vs_A} Systematics of $B(E1, 0^+_{gs} \rightarrow
1^-_1)$~strength in the lanthanide/rare-earth region. Figure is taken
from Ref.~\cite{ScheckHf}.}
\end{figure}

Interestingly, a long-standing theme investigated in $(\gamma,\gamma^{\prime})$ photon-scattering  experiments is the $E1$~strength of the $[2^+ \otimes 3^-]_{1^-}$ quadrupole-octupole
coupled (QOC) $1^-_1$ levels \cite{Kneissl96}. In particular, Kneissl,
Pitz, and coworkers established in stable nuclei comprehensive
systematics of these, in general, lowest-lying 1$^-_1$ levels
\cite{Fransen,Kneissl07}. The collective nature of these $1^-_1$ levels
is evidenced by their energy systematics
(Fig.~\ref{e1-a}); these systematics display a smooth behaviour, which
can be summarised as follows. At/near closed shells, the energy $E_{1^-}$
of the QOC 1$^{-}$ level corresponds nearly to the sum $\Sigma =
E_{2^+_1}+ E_{3^-_1}$ of the excitation energies of the first $2^+_1$
and $3^-_1$ levels, but decreases relative to $\Sigma$ with the onset
of quadrupole correlations. Transitional nuclei with ground-state
quadrupole correlations, but no well developed quadrupole deformation,
exhibit a near degeneracy of $1^-_1$ and $3^-_1$ levels. Once static
quadrupole deformation is present, levels 1$^-_1$  and
$1^-_2$ become the band-heads of the $K=0$ and $K=1$ octupole
bands. Furthermore, the $B(E1, 0^+ \rightarrow 1^-_1)$ strength
remains in the same order of magnitude over a wide range of nuclei
with varying underlying quadrupole deformation (Fig.~\ref{BE1_vs_A}).
In addition, in spherical nuclei it has been shown that the strength
of the  two-phonon creating/annihilating transition connecting this
$1^-$ state with the ground state scales with the $3^-_1 \rightarrow
2^+_1$ two-phonon exchanging transition \cite{Pietralla97}, whereas in
well-deformed nuclei the branching behaviour predicted by the Alaga
rules is observed \cite{ZilgesALAGA}. Interestingly, an enhanced E1
strength of up to $20 \times 10^{-3}$e$^{2}$fm$^{2}$ is noticeable
for semi-magic nuclei and prolate-deformed nuclei, which is almost an order of magnitude reduced in the transitional region. 

In spherical nuclei, for which the $2^+_1$ and $3^-_1$ levels are
interpreted as phonons, the QOC $1^-$ level is the low-spin member of
a quintuplet of negative-parity levels with spins ranging from
$J^{\pi} =1^-$ to $5^-$. However, candidate
levels for the full multiplet are proposed for only a few nuclei, see, e.g.,
Refs.~\cite{Wilhelm,JohnPaul}. While quadrupole deformation in
the ground state of the nuclear many-body quantum system is well
established, only a few candidates with enhanced octupole
correlations and possibly even deformation were proposed following
the observation of enhanced $B(E3, 0^+ \rightarrow 3^-_1)$ strength
in Coulomb-excitation experiments \cite{Butler,Gaffney,Wolle}.

\begin{figure}[t]
\resizebox{1.0\textwidth}{!}{\includegraphics{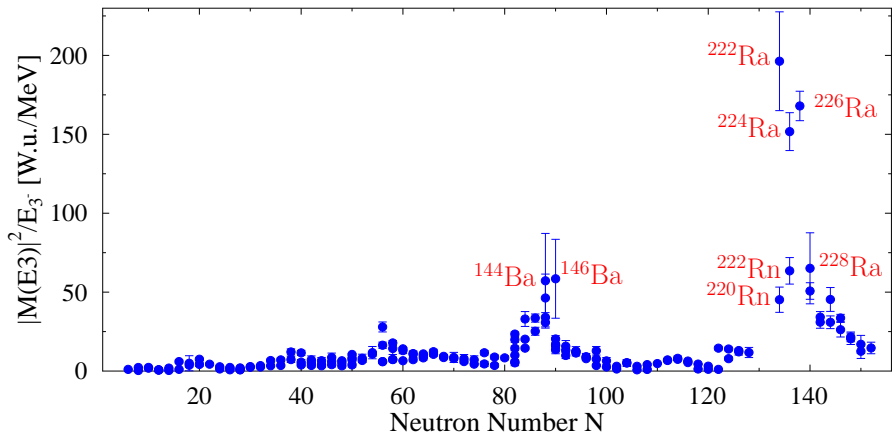}}
\caption{\label{Sumrule} Inverse energy-weighted $B(E3, 0^+
\rightarrow 3^-_1)$ strength as a function of the mass number $A$. The
systematics include nuclei for which the $B(E3)$ strength is
known \cite{Kibedi}. The recent data points for $^{144}$Ba
\cite{Bucher1}, $^{146}$Ba \cite{Bucher2}, $^{220,222}$Rn
\cite{Gaffney,Spagnoletti}, and $^{222,224,226,228}$Ra
\cite{Butler,Gaffney,Wolle} are highlighted.}
\end{figure}

Figure~\ref{Sumrule} shows the inverse energy-weighted
$B(E3)$ strengths. This quantity combines the two most relevant characteristics of octupole corre\-la\-tions/de\-for\-ma\-tions
and emphasises the special nature of the radium nuclei
$^{222,224,226}$Ra. Furthermore, for $^{228}$Th, indirect experimental evidence \cite{Majid} suggests enhanced octupole correlations. The
interplay of quadrupole deformation and octupole correlations, for
the above mentioned nuclei in the ground state, results in the
nucleus adopting a pear shape, in which the odd-electric (E1, E3, and E5) and even-magnetic (M2 and M4) moments
are present in the intrinsic reference frame. Indeed, quadrupole-octupole coupling is predicted to
enhance a possible nuclear Schiff moment
\cite{Spevak,Auerbach,Naftali,Zelevinsky,Dobacewski,Markus} and
magnetic quadrupole ($M2$) moment \cite{ViktorM2} caused by
CP-violating physics \cite{Musolf,Tim}.

The projection of the pear shape \cite{PeterWitek} from the intrinsic to the laboratory reference frame results in good-parity wave functions, that are linear combinations of the pear shape pointing to the left $\Psi_l$ and to the right $\Psi_r$. Assuming that $\Psi_l$ and $\Psi_r$ are orthogonal, the linear combinations are

\begin{equation}
\Psi^+ = \frac{1}{\sqrt{2}} \left( \Psi_l + \Psi_r \right),\quad\quad
\Psi^- = \frac{1}{\sqrt{2}} \left( \Psi_l - \Psi_r \right).
\end{equation}

\noindent The linear combination $\Psi^+$ is invariant under the
space-inversion $P$ operation. However, the $\Psi^-$ linear
combination inverts its sign and, therefore, has negative parity $\pi = -$. In an odd-mass nucleus, the coupling of the unpaired particle to these two states leads to the presence of parity
doublets, which are two levels with identical angular momentum but
opposite parity. Indeed, as shown in Table~\ref{PDs}, several
odd-mass nuclei in the region near $Z=88/90$ and $N=134/136$ exhibit
parity-doublet candidates. Clearly, the data in Table~\ref{PDs}
suffer from uncertainties concerning spin and, especially, parity
assignment, and unobserved upper partner levels. Nevertheless, at
present $^{227}$Ac is the nucleus with the lowest established
parity-doublet energy difference $\Delta E_{PD}$ of 27.4~keV.

In phenomenological estimates (see Refs.~\cite{Musolf,Tim} and references therein), the enhancement of the intrinsic nuclear Schiff moment, $\langle\hat{S}\rangle$, is predicted to scale with the quadrupole $\beta_2$ and the square of the octupole $\beta_3$ deformations, as well as inversely with the energy difference $\Delta
E_{PD}$ between the parity doublet partners
\begin{equation}
\langle\hat{S}\rangle \propto \frac{\beta_2 (\beta_3)^2}{\Delta E_{PD}}.
\end{equation}
\noindent The quadrupole-octupole coupling contributes with
the factor $\beta_2 \cdot \beta_3$ and the expected CP-violating
interaction with an additional $\beta_3$ factor. However, at present
it is not possible to disentangle the static and dynamic
contributions to the $\beta_3$ deformation parameter in a model-independent way. Nevertheless, self-consistent
calculations in well quadrupole-octupole deformed nuclei allow a determination of $\langle\hat{S}\rangle$ values directly, without
passing through the estimates of deformations $\beta_2$ and
$\beta_3$.

\begin{table}[t]
\begin{center}
\caption{\label{PDs} Data for selected odd-mass nuclei in the $A
\approx 224$ mass region exhibiting parity-doublet candidates. The table presents the isotope, its halflife $T_{1/2,gs}$, spin and parity of the
ground state $J_0^{\pi}$, energy difference to the lowest-lying
possible parity-doublet partner $\Delta E_{PD}$, and the lifetime
$T_{1/2,ul}$ of the upper level. Data are taken from the NNDC data
base \cite{NNDC} and Ref.~\cite{PTB}.}
 \begin{tabular}{ccccc}
\hline\noalign{\smallskip}
\hline\noalign{\smallskip}
 Nucleus & $T_{1/2,gs}$ & $J^{\pi}$ & $\Delta E_{PD}$ &$T_{1/2,ul}$\\
 & & & [keV]& [ns] \\
\hline\noalign{\smallskip}
$^{223}$Fr & 22.00(7)~m  & 3/2$^{(-)}$ & 134.48(4) & \\
$^{225}$Fr & 3.95(14)~m  & 3/2$^-$ & 142.59(3)  &\\
$^{227}$Fr & 2.47(3)~m   & 1/2$^+$ &  59.10(5)  &\\
$^{221}$Ra & 28(2)~s     & 5/2$^+$ & 103.61(11) &\\
$^{223}$Ra & 11.43(5)~d  & 3/2$^+$ &  50.128(9) & 0.63(7)\\
$^{225}$Ra & 14.9(2)~d   & 1/2$^+$ &  55.16(6)  &\\
$^{227}$Ra & 42.2(5)~m   & 3/2$^+$ &  90.034(2) & 0.254(9)\\
$^{223}$Ac & 2.10(5)~m   & (5/2$^-$) & 64.62(4) & $\leq$ 0.250\\
$^{225}$Ac & 9.920(3)~d  & (3/2$^-$) & 40.10(4) & 0.72(3)\\
$^{227}$Ac & 21.772(3)~y & 3/2$^-$ & 27.369(11) & 38.52(19)$^{a}$\\
$^{229}$Ac & 62.7(5)~m   & (3/2$^+$) & 104.3(4) & \\
$^{229}$Th & 7880(120)~y & 5/2$^+$ & 146.357(2) & \\
$^{231}$Th & 25.52(1)~h  & 5/2$^+$ & 185.718(2) & 1.07(8)\\
$^{229}$Pa & 1.50(5)~d   & (5/2$^+$) & 99.3(4) & \\
$^{231}$Pa & 32760(11)~y & 3/2$^-$ & 102.269(2) & $\leq$ 0.7\\
\noalign{\smallskip}\hline
\hline\noalign{\smallskip}
\end{tabular}
\end{center}
$^{a}$This value is taken from the most recent evaluation \cite{PTB}.
\end{table}

The above mentioned enhanced $B(E3)$~probabilities may indicate the
presence of a non-zero intrinsic $E3$ moment in the ground state of Ra isotopes and, consequently, a possibility of the additional non-zero intrinsic $E1$ and $E5$ as well as $M2$ and $M4$ moments. We can thus expect that a CP-violating interaction may induce enhanced
values of the corresponding symmetry-violating laboratory moments.

\section{Experimental motivation for studying $^{227}$Ac}

Given the high charge number of $Z = 89$ and low energy of the
transition connecting the parity-partner levels in $^{227}$Ac, and
considering that the conversion coefficients (CCs) exhibit a strong
multipolarity dependence, a measurement of the CCs appear to be the obvious way to search for CP-violating physics. If parity was no longer a good
quantum number, the $E1$ transition connecting the parity-doublet
partners would contain a $M1$ component. However, it can be
assumed that the P-/T-odd effect scales as a $10^{-7}$
contribution of the weak interaction to the nuclear force. Hence, any
signal of such physics will be well below the associated
uncertainty in the calculation of CCs. For the 27.4-keV
$E1$~transition in $^{227}$Ac, the value of the CC is 
$\alpha_C = 3.54(5)$ \cite{BRICC}, which still allows for $\approx
22$~\% of all decays to proceed via $\gamma$-ray emission.

The low $\gamma$-ray energy, and hence the low momentum
transfer in the emission and absorption process, enables the nuclear
photonics technique of recoil-free resonant absorption
(known as M\"ossbauer spectroscopy \cite{Moessbauer,Guetlich1,Guetlich2}) to be used as a method to investigate the $E1$ transition in question and, subsequently, to pin down the properties of the two parity-doublet partners. M\"ossbauer spectroscopy exploits the fact that, for a nucleus embedded in a crystal lattice, there is a probability that the recoil momentum transfer in the emission and absorption of a $\gamma$~ray is absorbed by the entire crystal. If both processes are recoil-free, the theoretical achievable energy-resolution $\Delta E \approx 2 \cdot \Gamma_{\gamma}$ is limited only by the natural line width $\Gamma$, which for $^{227}$Ac is $\Gamma_{\gamma} = \frac{\hbar \ln(2)}{T_{1/2,ul}} = 11.8(1) \times 10^{-9}$~eV.

Other advantages of studying $^{227}$Ac are its presence in the
decay chain of $^{235}$U and a comparably long half-life ($T_{1/2} \approx 21.8$~years \cite{NNDC}). These may allow a chemical separation of a
sufficient amount of the target material to manufacture an absorber target, whose lattice would be tailored to the specific requirements of M\"ossbauer spectroscopy.

M\"ossbauer spectroscopy can be performed in fluorescence or
absorption. The latter, for which the detector is situated in the
extension of the emitter-absorber line, has the advantage that the
detector can be retracted from the absorber/emitter samples. This
prevents pile-up events due to a too high detector count rate. Given the
relatively small expected size of the radioactive samples, the
resulting reduction of the solid angle would not adversely impact the
count rate of good events.

A further benefit of $^{227}$Ac is that the upper parity-doublet
level is strongly populated following either the $\alpha$~decay of
$^{231}$Pa ($T_{1/2}= 32760(110)$~years) \cite{Teoh} or the
$^{227}$Ra $\beta$ decay ($T_{1/2} = 42.2(5)$~min) \cite{Lourens}.
While practical considerations involving the lifetime of $^{231}$Pa
favour the population via $\alpha$ decay, the recoil experienced by
the $^{227}$Ac daughter might be devastating in terms of a
well-defined position of the emitting $^{227}$Ac nucleus within the lattice
structure and, therefore, local field distribution.

\section{CP-violating moments and M\"ossbauer spectroscopy}

Traditionally, three effects can be observed in M\"ossbauer
spectroscopy, namely, isomer shift, qua\-dru\-pole splitting, and
nuclear Zeeman effect. Each of them slightly shifts the
energies of the involved levels and, consequently, alters the energy
of the emitted and/or absorbed $\gamma$~rays. The resulting energy
difference can be compensated by mounting the emitter or absorber
source on a drive inducing a velocity-dependent Doppler shift. For
example, for $^{227}$Ac an energy shift of $11.74 \times 10^{-9}$~eV
corresponds to a sample velocity of $1$~cm/s.

\begin{figure}[t]
\resizebox{1.0\textwidth}{!}{\includegraphics{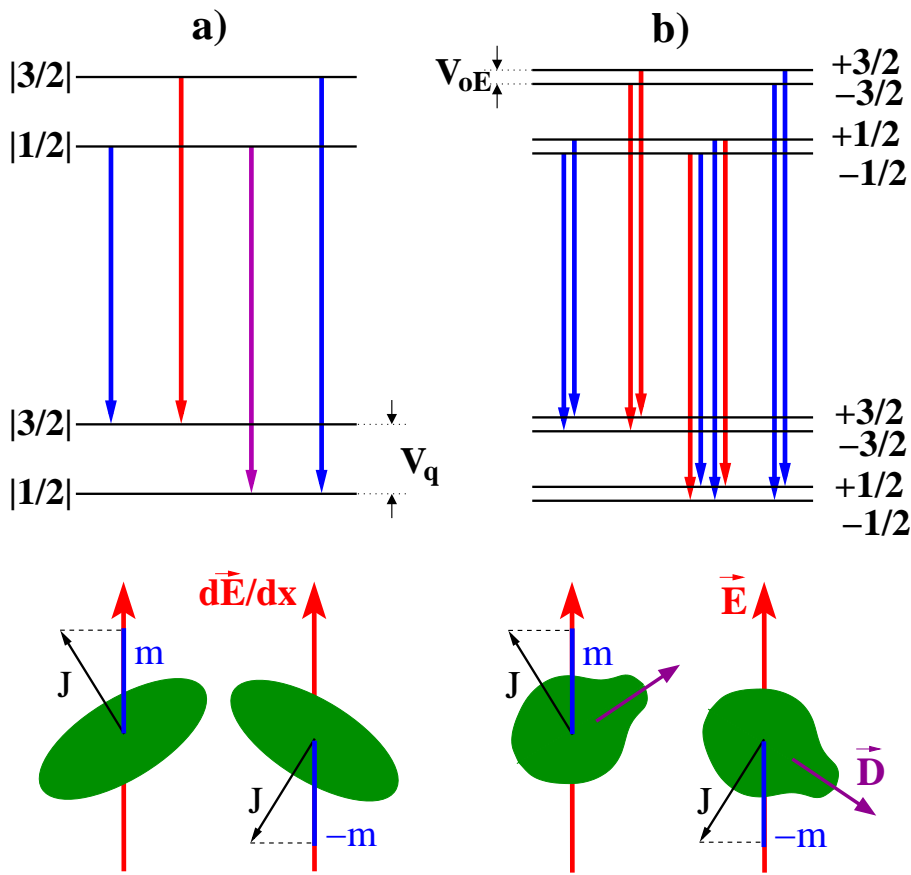}}
\caption{\label{QuadrupoleSplitting} Part~a) shows the level scheme
of a nucleus with a $J^{\pi}=3/2^{\pm}$ parity doublet under the
influence of quadrupole splitting. Part~b) includes the influence of the interaction $V_{oE}$ of an
$EL$ odd-electric moment with the corresponding $L^{th}$ derivative
of the electric potential of the crystal lattice at the nuclear
coordinates. This CP-violating physics lifts the $\vert m \vert$
degeneracy of the time-reversal orbits with magnetic quantum numbers
$m$ and $-m$. For a detailed discussion see text.}
\end{figure}

The {\bf isomer shift} corresponds to a slight shift of the Cou\-lomb
energy of a nuclear level due to the electron density and,
consequently, electrostatic potential at the nuclear coordinates; it
is observed, if emitter and absorber nuclei are embedded in different
crystal lattices. The second effect of {\bf quadrupole splitting} is
observed if, due to the crystal composition at the coordinates of
the absorbing nucleus, an electric-field gradient is present. This
situation is shown in Fig.~\ref{QuadrupoleSplitting}~a) for a nucleus
with a $J^{\pi}=3/2^{\pm}$ parity doublet. The reflection symmetry of
the quadrupole shape results in a $\vert m \vert$ degeneracy of the time-reversal
$m$ and $-m$ orbits. For identical quadrupole moments of the
parity-doublet partners, and, therefore, identical splitting, the
spectrum would contain three transitions, with $\Delta\vert m\vert= \pm 1$
shifted to lower/higher energy and the $\Delta \vert m\vert= 0$ transitions at
the unperturbed transition energy. However, due to the parity-doublet partners being linear combinations, it can be expected that their quadrupole moments differ and the transition energies of the two $\Delta\vert m\vert= 0$ lines will be different. Hence, M\"ossbauer spectroscopy represents a sensitive test to extract at least the relative difference of the quadrupole moments or, since the quadrupole moment of the $J^{\pi} = 3/2^-$ ground state is known, the quadrupole moment of the upper partner level.  For such a test it is of benefit that line-shape parameters, especially the Full-Width at Half Maximum, can be fitted to the two $\Delta\vert m\vert =\pm 1$ transitions and even a small widening of the central line consisting of the two $\Delta\vert m\vert= 0$ transitions would allow an extraction of the difference.

Far more intriguing is that, if the investigated nucleus is truly reflection asymmetric, any residual interaction of an odd-electric moment with a higher moment of the multipole expansion of the lattice charge
distribution at the nuclear coordinates (e.g., $E1$~moment with the
electric field or the $E3$~moment with the curvature of the electric
field) would lift the $\vert m\vert$ degeneracy. Again, given that $EL$ moments of the two physical states represented by the linear combinations differ,  
a different interaction $V_{oE}$ of $EL$ odd-electric moment and $L^{th}$ multipole order of the lattice charge distribution can be anticipated for each doublet partner level. Consequently, a split in the transitions will be observed. With the examption of the $\Delta m = \pm 1$ transitions of the $\Delta\vert m\vert =0$, namely the $m_{\text{upper}} = \pm 1/2 \rightarrow m_{\text{lower}} = \mp 1/2$ transitions, for all transitions the splitting results in the same energy shift. Only these $\Delta m = \pm 1$ transitions experience a stronger shift and the $\Delta \vert m\vert =0$ transition will, in addition to the quadrupole splitting exhibit a further splitting into a quadruplet. In the quadruplet the two satellite lines are expected to show half the intensity of the lesser shifted $\Delta\vert m\vert =0$, $\Delta m=0$ transitions. It is helpful that the two $\Delta\vert m\vert= \pm 1$ lines are not affected in the same way and allow a fit of the peak shape. Consequently, even if $V_{oE}$ is less than
$\Delta E$, an alteration of the central line's peak shape yields
evidence for CP-violating physics possibly even an order of magnitude
below the experimental resolution $\Delta E$. Concerning the
interaction of a nuclear $E1$ moment and the electric field at the
nuclear coordinates, it must be mentioned that, in a given lattice
structure, the nucleus will position itself at coordinates for which
the net electric field vanishes. Therefore, the $E3$ moment should be
the lowest odd-electric moment contributing. Eventually, the use of a
piezzo-electric lattice will allow access to the $E1$ moment.

Finally, the combination of the sensitivity of M\"ossbauer
spectroscopy and the magnetic field at the nuclear coordinates in a
lattice allow the observation of the {\bf nuclear Zeeman effect}.
If the magnetic moments of the parity-doublet partners are identical,
the three-line pattern of the normal Zeeman effect will be observed.
However, due to the two parity partners being linear combinations,
their magnetic properties differ and the pattern of the anomalous
Zeeman effect is expected. Here, for a $J=3/2$ parity doublet, ten transitions can be observed. However, these lines will exhibit a centroid symmetry. Given it is present, the interaction of a CP-forbidden magnetic quadrupole $M2$ moment \cite{ViktorM2} with the magnetic-field gradient perturbs the expected pattern. Since the additional interaction is direction-dependent, it can be expected that the otherwise $m$-in\-de\-pen\-dent Zeeman splitting between two levels with $m=0$ and $m \pm 1$ receives an additional $m$-dependent term. For such a $m$-dependent splitting, the centroid-symmetric pattern will be disturbed.

To extract a quantitative result for a possible CP-violating
interaction, knowledge of the electric and magnetic field
distribution, specifically the field values and higher derivatives at
the position of the nucleus, is required. These values can be
calculated for specific materials using modern density functional
theory \cite{Martin}, and this has already been successfully used to
obtain electric field gradient and hyperfine fields for a variety of
materials (see, e.g., \cite{Bluegel,Blaha1,Blaha2}).

\section{Summary and Outlook}

To summarise, in this contribution, we proposed M\"ossbauer spectroscopy
of the octupole-correlated $^{227}$Ac nucleus as a new avenue to search for
CP-violating physics. This work provides first thoughts in relation to the way in which a residual interaction associated with an odd-electric or even-magnetic moment alters the effects of qua\-drupole splitting or the nuclear Zeeman effect. In a long-term future outlook, this project would benefit from a sufficiently mono-energetic source with $\Delta E_{\gamma} \ll
\Gamma$ of a fully-polarised $\gamma$-ray beam. This would allow the
elimination of effects associated with recoils in the radioactive decays
and in addition the selective population of $m$ substates can be used to test the involved $m$~substates.

\bmhead{Acknowledgments}

We acknowledge financial support by the UK-STFC Grant
Nos.~ST/P005101/1, ST/P003885/1, and~ST/V001035/1, by the Polish
National Science Centre under Contract No.~2018/31/B/ST2/02220, and
by a Leverhulme Trust Research Project Grant. We acknowledge the
CSC-IT Center for Science Ltd., Finland, for the allocation of
computational resources. This project was partly undertaken on the
Viking Cluster, which is a high performance computing facility provided
by the University of York. We are grateful for computational support
from the University of York High Performance Computing service,
Viking and the Research Computing team.


\begin{thebibliography}{00}

\bibitem{Bucher1} B.~Bucher, S.~Zhu, C.~Y.~Wu, R.~V.~F.~Janssens, D.~Cline, A.~B.~Hayes, M.~Albers, A.~D.~Ayangeakaa, P.~A.~Butler, C.~M.~Campbell, M.~P.~Carpenter, C.~J.~Chiara, J.~A.~Clark, H.~L.~Crawford, M.~Cromaz, H.~M.~David, C.~Dickerson, E.~T.~Gregor, J.~Harker, C.~R.~Hoffman, B.~P.~Kay, F.~G.~Kondev, A.~Korichi, T.~Lauritsen, A.~O.~Macchiavelli, R.~C.~Pardo, A.~Richard, M.~A.~Riley, G.~Savard, M.~Scheck, D.~Seweryniak, M.~K.~Smith, R.~Vondrasek, A.~Wiens, Phys.~Rev.~Lett. {\bf 116}, 112503 (2016)
\bibitem{Bucher2} B.~Bucher, S.~Zhu, C.~Y.~Wu, R.~V.~F.~Janssens, R.~N.~Bernard, L.~M.~Robledo, T.~R.~Rodriguez, D.~Cline, A.~B.~Hayes, A.~D.~Ayangeakaa, M.~Q.~Buckner, C.~M.~Campbell, M.~P.~Carpenter, J.~A.~Clark, H.~L.~Crawford, H.~M.~David, C.~Dickerson, J.~Harker, C.~R.~Hoffman, B.~P.~Kay, F.~G.~Kondev, T.~Lauritsen, A.~O.~Macchiavelli, R.~C.~Pardo, G.~Savard, D.~Seweryniak, R.~Vondrasek, Phys.~Rev.~Lett. {\bf 118}, 152504 (2017)
\bibitem{Spagnoletti} P.~Spagnoletti, P.~A.~Butler, L.~P.~Gaffney, K.~Abrahams, M.~Bowry, J.~Cederkall, T.~Chupp, G.~de~Angelis, H.~De~Witte, P.~E.~Garrett, A.~Goldkuhle, C.~Henrich, A.~Illana, K.~Johnston, D.~T.~Joss, J.~M.~Keatings, N.~A.~Kelly, M.~Komorowska, J.~Konki, T.~Kr\"oll, M.~Lozano, B.~S.~Nara~Singh, D.~O'Donnell, J.~Ojala, R.~D.~Page, L.~G.~Pedersen, C.~Raison, P.~Reiter, J.~A.~Rodriguez, D.~Rosiak, S.~Rothe, M.~Scheck, M.~Seidlitz, T.~M.~Shneidman, B.~Siebeck, J.~Sinclair, J.~F.~Smith, M.~Stryjczyk, P.~Van~Duppen, S.~Vinals, V.~Virtanen, K.~Wrzosek-Lipska, N.~Warr, M.~Zielinska, Phys.~Rev.~C {\bf 105}, 024323 (2022)
\bibitem{Butler} P.~A.~Butler, L.~P.~Gaffney, P.~Spagnoletti, K.~Abrahams, M.~Bowry, J.~Cederkall, G.~de~Angelis, H.~De~Witte, P.~E.~Garrett, A.~Goldkuhle, C.~Henrich, A.~Illana, K.~Johnston, D.~T.~Joss, J.~M.~Keatings, N.~A.~Kelly, M.~Komorowska, J.~Konki, T.~Kr\"oll, M.~Lozano, B.~S.~Nara~Singh, D.~O'Donnell, J.~Ojala, R.~D.~Page, L.~G.~Pedersen, C.~Raison, P.~Reiter, J.~A.~Rodriguez, D.~Rosiak, S.~Rothe, M.~Scheck, M.~Seidlitz, T.~M.~Shneidman, B.~Siebeck, J.~Sinclair, J.~F.~Smith, M.~Stryjczyk, P.~Van~Duppen, S.~Vinals, V.~Virtanen, N.~Warr, K.~Wrzosek-Lipska, M.~Zielinska, Phys.Rev.Lett. {\bf 124}, 042503 (2020)
\bibitem{Gaffney} L.~P.~Gaffney, P.~A.~Butler, M.~Scheck, A.~B.~Hayes, F.~Wenander, M.~Albers, B.~Bastin, C.~Bauer, A.~Blazhev, S.~B\"onig, N.~Bree, J.~Cederkall, T.~Chupp, D.~Cline, T.~E.~Cocolios, T.~Davinson, H.~De~Witte, J.~Diriken, T.~Grahn, A.~Herzan, M.~Huyse, D.~G.~Jenkins, D.~T.~Joss, N.~Kesteloot, J.~Konki, M.~Kowalczyk, Th.~Kr\"oll, E.~Kwan, R.~Lutter, K.~Moschner, P.~Napiorkowski, J.~Pakarinen, M.~Pfeiffer, D.~Radeck, P.~Reiter, K.~Reynders, S.~V.~Rigby, L.~M.~Robledo, M.~R\"udigier, S.~Sambi, M.~Seidlitz, B.~Siebeck, T.~Stora, P.~Thoele, P.~Van~Duppen, M.~J.~Vermeulen, M.~von~Schmid, D.~Voulot, N.~Warr, K.~Wimmer, K.~Wrzosek-Lipska, C.~Y.~Wu, M.~Zielinska, Nature {\bf 497}, 199 (2013)
\bibitem{Wolle} H.~J.~Wollersheim, H.~Emling, H.~Grein, R.~Kulessa, R.~S.~Simon, C.~Fleischmann, J.~de~Boer, E.~Hauber, C.~Lauterbach, C.~Schandera, P.~A.~Butler, T.~Czosnyka, Nucl.~Phys.~{\bf A556}, 261 (1993)
\bibitem{Spevak} V.~Spevak and N.~Auerbach, Phys.~Lett. {\bf B359}, 254 (1995)
\bibitem{Auerbach} N.~Auerbach, V.~V.~Flambaum, and V.~Spevak,  Phys.~Rev.~Lett. {\bf 76}, 4316 (1996)
\bibitem{Naftali}  N.~Auerbach, V.~F.~Dmitriev, V.~V.~Flambaum, A.~Lisetskiy, R.~A.~Senkov, and V.~G.~Zelevinsky, Phys.~Rev.~C {\bf 74}, 025502 (2006)
\bibitem{Zelevinsky} V.~G.~Zelevinsky, A.~Volya, and N.~Auerbach, Phys.~Rev.~C
{\bf 78}, 014310 (2008)
\bibitem{Dobacewski} J.~Dobaczewski and J.~Engel, Phys.~Rev.~Lett. {\bf 94}, 232502 (2005)
\bibitem{Markus} J.~Dobaczewski, J.~Engel, M.~Kortelainen, and P.~Becker,
Phys.~Rev.~Lett. {\bf 121}, 232501 (2018)
\bibitem{ViktorM2} V.~V.~Flambaum and A.~J.~Mansour,  Phys. Rev. C {\bf 105}, 065503 (2022)
\bibitem{Kneissl96} U.~Kneissl, H.~H.~Pitz, and A.~Zilges, Prog.~Part.~Nucl.~Phys {\bf 37}, 349 (1996)
\bibitem{Fransen} C.~Fransen, O.~Beck, P.~von~Brentano, T.~Eckert, R.-D.~Herzberg, U.~Kneissl, H.~Maser, A.~Nord, N.~Pietralla, H.~H.~Pitz, and A.~Zilges, Phys.~Rev.~C {\bf 57}, 129 (1998)
\bibitem{Kneissl07} U.~Kneissl, N.~Pietralla, and A.~Zilges, J.~Phys.~G {\bf 32}, R217 (2006)
\bibitem{Pietralla97} N.~Pietralla, Phys.~Rev.~C {\bf 59}, 2941 (1997)
\bibitem{ZilgesALAGA} A.~Zilges, P.~von~Brentano, A.~Richter, R.~D.~Heil, U.~Kneissl, H.~H.~Pitz, C.~Wesselborg, Phys.~Rev.~C {\bf 42}, 1945 (1990)
\bibitem{ScheckHf} M.~Scheck, D.~Belic, P.~von~Brentano, J.~J.~Carroll, C.~Fransen, A.~Gade, H.~von~Garrel, U.~Kneissl, C.~Kohstall, A.~Linnemann, N.~Pietralla, H.~H.~Pitz, F.~Stedile, R.~Toman, and V.~Werner,  Phys.~Rev.~C {\bf 67}, 064313 (2003)
\bibitem{Wilhelm} M.~Wilhelm, E.~Radermacher, A.~Zilges, and P.~von~Brentano,  Phys.\@ Rev.~C {\bf 54}, R449 (1996)
\bibitem{JohnPaul} P.~E.~Garrett and J.~L.~Wood, J.\@ Phys.~G {\bf 37}, 064028 (2010)
\bibitem{Majid} M.~M.~R.~Chishti, D.~O'Donnell, G.~Battaglia, M.~Bowry, D.~A.~Jaroszynski, B.~S.~Nara~Singh, M.~Scheck, P.~Spagnoletti, J.~F.~Smith, Nature Physics {\bf 16}, 853 (2020)
\bibitem{Musolf} J.~Engel, M.~J.~Ramsey-Musolf, U.~van~Kolck,  Prog.\@ Part.\@ Nucl.\@ Phys.\@ {\bf 71}, 21 (2013)
\bibitem{Tim} T.~Chupp and M.~Ramsey-Musolf, Phys.\@ Rev.~C {\bf 91}, 035502 (2015)
\bibitem{Kibedi} T.~Kib\'{e}di and R.~H.~Spears, At.\@ Data and Nucl.\@ Data Tables
{\bf 80}, 35 (2002)
\bibitem{PeterWitek} P.~A.~Butler and W.~Nazarewicz, Rev.~Mod.~Phys. {\bf 68}, 349 (1996)
\bibitem{NNDC} http:www.nndc.bnl.gov (accessed 12.10.2022)
\bibitem{PTB} M.~P.~Takacs and K.~Kossert,  Appl.\@ Radiat.\@ Isot. {\bf 176}, 109858 (2021)
\bibitem{BRICC}  T.~Kibedi, T.~W.~Burrows, M.~B.~Trzhaskovskaya, P.~M.~Davidson, C.~W.~Nestor Jr., Nucl.\@ Instr.\@ and Meth.~A {\bf 589}, 202 (2008) 202
\bibitem{Moessbauer} R.~L.~M\"ossbauer. Z.~Physik {\bf 151}, 124 (1958)
\bibitem{Guetlich1} P.~G\"utlich, E.~Bill, and A.~X.~Trautwein, {\it M\"ossbauer Spectroscopy and Transition Metal Chemistry}, Springer (2011)
\bibitem{Guetlich2} P.~G\"utlich and C.~Schr\"oder, {\it M\"ossbauer Spectroscopy, Methods in Physical Chemistry,}, Wiley-VCH, pp.~351 (2012)
\bibitem{Teoh} W.~Teoh, R.~D.~Connor, and R.~H.~Betts, Nucl.\@ Phys.~{A319}, 122 (1979)
\bibitem{Lourens} W.~Lourens, B.~O.~Ten~Brink, and A.~H.~Wapstra, Nucl.\@ Phys.~{\bf A179}, 337 (1971)
\bibitem{Martin} R.~Martin, {\it Electronic Structure: Basic Theory and Practical Methods}, Cambridge University Press (2004).
\bibitem{Bluegel} S. Bl\"ugel, H. Akai, R. Zeller, and P. Dederichs, Phys.\@ Rev.\@ B {\bf 35}, 3271 (1987)
\bibitem{Blaha1} P.~Blaha, P.~Dufek, and K.~Schwarz, Hyperfine Interactions {\bf 95}, 257 (1995)
\bibitem{Blaha2} P. Blaha, J.\@ Phys.\@: Conf.\@ Series {\bf 217}, 012009 (2009)
\end{thebibliography}


\end{document}